# Dynamical-mean-field cluster-correlation expansion of NV− spin-echo decoherence in a realistic substitutional-nitrogen bath

Chikara Shinei[1],[2]

[1] *Department of Applied Physics, Institute of Pure and Applied Sciences, University of Tsukuba, Tsukuba, Ibaraki 305-8573, Japan*

[2] *Japanese-French Laboratory for Semiconductor Physics and Technology J-FAST, CNRS, University Grenoble Alpes, Grenoble INP, University of Tsukuba, Tsukuba, Japan*

*shinei.chikara.fb@u.tsukuba.ac.jp*

## Abstract

The Hahn-echo time $T_2$ of nitrogen-vacancy (NV−) ensembles in nitrogen-rich diamond follows the empirical relation $T_2 \cdot [N] = 160$ μs·ppm, and this relation does not depend on the magnetic field between 2 and 30 mT. Cluster-correlation-expansion (CCE) calculations solve the spin dynamics exactly within small groups of spins. When they treat the Overhauser field of the surrounding P1 centres, that is, the field that these electron spins produce, as frozen, they overestimate $T_2$ at high field by up to a factor of two. This excess was recently attributed to additional dark defects. We show that the excess comes mostly from the frozen field. In our model the P1 centres outside each cluster flip-flop in pairs: two centres exchange their spin states and conserve their total $S_z$. Kinetic Monte Carlo generates these events at golden-rule rates, which follow from the dipolar Hamiltonian without adjustable parameters, and each cluster evolves in the resulting time-dependent field. This dynamical mean field lies between the static-mean-field and mean-field-free limits. For a $^{14}$N P1 bath it reduces $T_2 \cdot [N]$ at 30 mT from 264 ± 16 to 188 ± 12 μs·ppm and gives 167 ± 11 μs·ppm at 10 mT, 1.7 and 0.4 standard deviations from experiment, with no dark spins. When each centre jumps independently, $T_2 \cdot [N]$ is 147 ± 9 μs·ppm at 30 mT; the correlation of the flip-flop partners thus lengthens $T_2$ by 20–30 % at high field. The Overhauser field is neither frozen nor averaged out by fast motion: 68–83 % of its autocorrelation survives 200 μs. At 2.5 mT, where the P1 electron and nucleus hybridise, $T_2$ stays below experiment for any choice of rates. Part of the field dependence therefore remains: $T_2$ rises by a factor 1.69 ± 0.07 instead of 2.15 ± 0.08 from 2.5 to 30 mT.

## I. INTRODUCTION

Ensembles of negatively charged nitrogen-vacancy (NV−) centres are the workhorse of solid-state quantum magnetometry [1]. In the nitrogen-rich diamond of ensemble sensors, the electron spins of substitutional nitrogen (P1 centres) form a spin bath, and this bath limits the coherence. Both the dephasing time $T_2^*$ and the Hahn-echo time $T_2$ scale inversely with the nitrogen concentration [N]. Bauch et al. measured 25 diamonds and established the relation $T_2 = (160 \pm 12)$ μs·ppm / [N] [2]. Within about 10 %, this relation does not depend on the bias field between 2 and 30 mT. It is therefore the natural benchmark for any microscopic theory of NV decoherence in a P1 bath.

The echo time is harder to compute than the dephasing time, and the reason is instructive. Bauch et al. accompanied their measurements with a semiclassical model [2]. In this model the P1 bath acts on the NV centre as Gaussian Ornstein–Uhlenbeck noise, a random noise with rms amplitude $\Delta$ and correlation time $\tau_c$. The positions of the P1 centres alone fix the amplitude, which gives $T_2^* = 1/\Delta_{ens}$. The correlation time comes from the flip-flop rates of independent P1 pairs; in a flip-flop, one spin flips up while its partner flips down. The model treats each pair as a two-level system with a detuning $\delta$ (an energy mismatch). It broadens this detuning by a common dephasing rate $\Gamma_d$, which it sets to the dipolar linewidth of the bath. We extended this model with the Jahn-Teller-resolved P1 hyperfine tensor [3]. It then gives $T_2^*\cdot[N]$ = 11.2 μs·ppm, within 17 % of the measured 9.6 μs·ppm, but $T_2\cdot[N]$ = 44 μs·ppm, a factor 3.6 below 160 μs·ppm. Bauch et al. themselves noted that their $T_2$ is less accurate than $T_2^*$. They attributed this to the many-body dynamics of the bath, which a stochastic mean-field model cannot capture [2]. The asymmetry has a simple origin. $T_2^*$ probes only the static distribution of the bath field, whereas $T_2 = (2\tau_c/\Delta^2_{ens})^{1/3}$ [2] probes its dynamics. Because of the cube root, a factor 3.6 in $T_2$ corresponds to a factor 48 in $\tau_c$: the semiclassical bath fluctuates some fifty times too fast. The likely cause is the treatment of the local field that detunes each pair. The model sets the rate of change $\Gamma_d$ of this field to the full dipolar linewidth. As we show below, however, the field stays largely static during the echo. The model thus turns a nearly static field into fast homogeneous dephasing. This broadens every pair resonance, and detuned pairs then flip-flop through the Lorentzian tail of the rate expression. The central question for $T_2$ is therefore how to treat the local field of the surrounding bath: as fast noise, as a frozen detuning, or not at all. A many-body method that treats the pair dynamics exactly can address this question.

The cluster-correlation expansion (CCE) [4,5] is such a method. It treats the bath exactly within small groups of spins (clusters) and builds the coherence from their contributions. Several studies have applied it to dilute baths of dipole-coupled electron spins [6–8]. These calculations need an assumption about the spins outside each cluster, and their results depend sharply on it. In one choice the outer spins are ignored; the bath is then fully mixed, i.e. in a random, unpolarised state. Every pair of P1 centres flip-flops as if isolated, and $T_2$ comes out well below experiment. In the other choice each outer spin is put in a random state. Its contribution to the Overhauser field, the magnetic field that the bath spins produce at a given spin, then enters the cluster Hamiltonian. This field detunes the pair flip-flops, and $T_2$ grows. Park et al. recently carried out such a calculation for P1 baths [6]. Their Hahn-echo times, computed at 50 mT, are about twice the values that they measured at 28–30 mT in samples with 0.8 and 13 ppm of P1. They reproduced the measurements with further paramagnetic defects: unidentified dark electron spins (X spins) of the kind discussed for bulk diamond [1]. They chose the concentrations of these spins, 0.4 and 6 ppm, to match the data; these values are about half the P1 concentration in both samples.

Before more defects enter the bath, it is worth asking what the measurements themselves indicate. The relation $T_2\cdot[N]$ = 160 μs·ppm was calibrated against independently measured nitrogen concentrations [2]. Yamamoto et al. recently tested it at the opposite end of the concentration range [9]. They studied single NV centres in isotopically engineered diamond. Their dynamical-decoupling coherence time, measured with trains of n refocusing pulses, grows as $n^{0.67(2)}$; this 2/3 law is characteristic of an electron-spin bath. From that regime they extrapolated a Hahn-echo time of 16 ms. Through the 160 μs·ppm relation, this time gives $[N] \approx 10$ ppb. This value agrees with their independent estimate from secondary-ion mass spectrometry (below the 20 ppb detection limit) and from the growth conditions. The Hahn-echo times that Park et al. measured, 163.3 μs at 0.8 ppm and 10.5 μs at 13 ppm [6], correspond to 131 and 137 μs·ppm. They also lie close to the same relation.

Direct tests have also asked whether unidentified spins are needed. Shinei et al. measured $T_2$ in HPHT and CVD crystals that contained $NV^-$ and $NV^0$ centres at concentrations comparable to that of P1, and they quantified all three [10]. Over 1–50 ppm, $T_2$ was inversely proportional to the sum of the three concentrations. $T_2$ did not change when electron irradiation and annealing converted P1 into NV centres at a constant sum. It was also insensitive to $NVH^-$, a defect common in CVD diamond. The $NV^-$ centre decoheres the ensemble with a coefficient of 89 μs·ppm, about 1.8 times more strongly per ppm than P1.

The samples of Park et al. contain 0.3 and 4.5 ppm of NV centres [6]. These values are about 40 % of the P1 concentration and close to the dark-spin concentrations that they fitted. With the empirical coefficients above, P1 alone would give 200 and 12.3 μs, and P1 together with the NV centres would give 120 and 7.6 μs. The measured 163.3 and 10.5 μs lie between the two. The identified nitrogen-related centres thus account for the measured echo times, and no additional unidentified spins are needed. Over five orders of magnitude in nitrogen concentration, $T_2$ follows the nitrogen content as a P1-dominated bath would, and no sample-dependent additional bath appears. This evidence cannot exclude defects whose concentration is proportional to [N], because the empirical constant would absorb them. It does suggest, however, that a P1-only calculation that overshoots 160 μs·ppm by a factor of two should be examined before the excess is assigned to additional defects.

One candidate for such an artefact is the treatment of the field itself. The Overhauser field on a pair comes from P1 centres that themselves flip-flop with their own neighbours, on the same time scale as the NV decoherence. A field that is sampled once and then frozen for the whole echo therefore detunes the pair too effectively. Indeed, in a static-mean-field CCE, $T_2$ grows by a factor of about two between 2.5 and 30 mT (Sec. III). This contradicts the field independence that Bauch et al. reported [2]. A mean-field-free CCE, in contrast, stays well below experiment at all fields. Both observations point to the frozen field as the culprit, but neither limit can tell where between the two the true bath lies.

Earlier work has treated the outer bath in two ways. CCE studies keep the exact dynamics inside each cluster, but they fix the state of the outer spins: the outer spins are either fully mixed [4,5] or put in random product states, whose static field then enters each cluster [6,11,12]. Stochastic models take the opposite approach. They describe the bath as classical noise on the central spin, for example Ornstein–Uhlenbeck noise [2,13,14] or the random processes of spectral-diffusion theory [15–17]. These models do let the field change in time. However, they take its amplitude and correlation time as input or from simplified rate estimates, and they lose the exact pair dynamics that the CCE provides. Neither approach derives the motion of the outer field from the bath Hamiltonian and also keeps the exact cluster dynamics.

Here we close that gap with a dynamical mean field, an outer field that changes in time. The P1 centres outside a cluster evolve by the same process that makes the field fluctuate: pairs of neighbouring centres flip-flop. In each flip-flop the two centres exchange polarisation between their electron–nuclear eigenstates and conserve their total $S_z$. We do not fit the flip-flop rates. They follow from the dipolar matrix elements of the same Hamiltonian through Fermi's golden rule, evaluated at the instantaneous detuning of each pair, in the spirit of the stochastic theories of spectral diffusion [15–17]. Each cluster then evolves exactly in the resulting time-dependent Overhauser field, and we average the echo over field trajectories. The method differs from the earlier treatments in three respects. First, the rates of the outer-bath dynamics follow from the bath Hamiltonian alone, with no adjustable parameters. Second, the two partners of each flip-flop change together and conserve their total $S_z$, as in the real bath. Third, the method contains both standard CCE treatments as exact limits: zero rates freeze the field, and infinite rates average it to zero. Our code reproduces both limits as computed with PyCCE [12]. We also compare the pair flip-flops with a simpler variant. In this variant every centre jumps on its own at its summed golden-rule rate. The variant costs less, but it ignores that the two partners of a flip-flop change together. Figure 1 contrasts the three treatments. We find three results. The fluctuating field removes most of the factor-of-two excess at 30 mT without additional defects. The pair correlation changes $T_2$ by 20–30 %. At low field the P1-only picture remains incomplete.

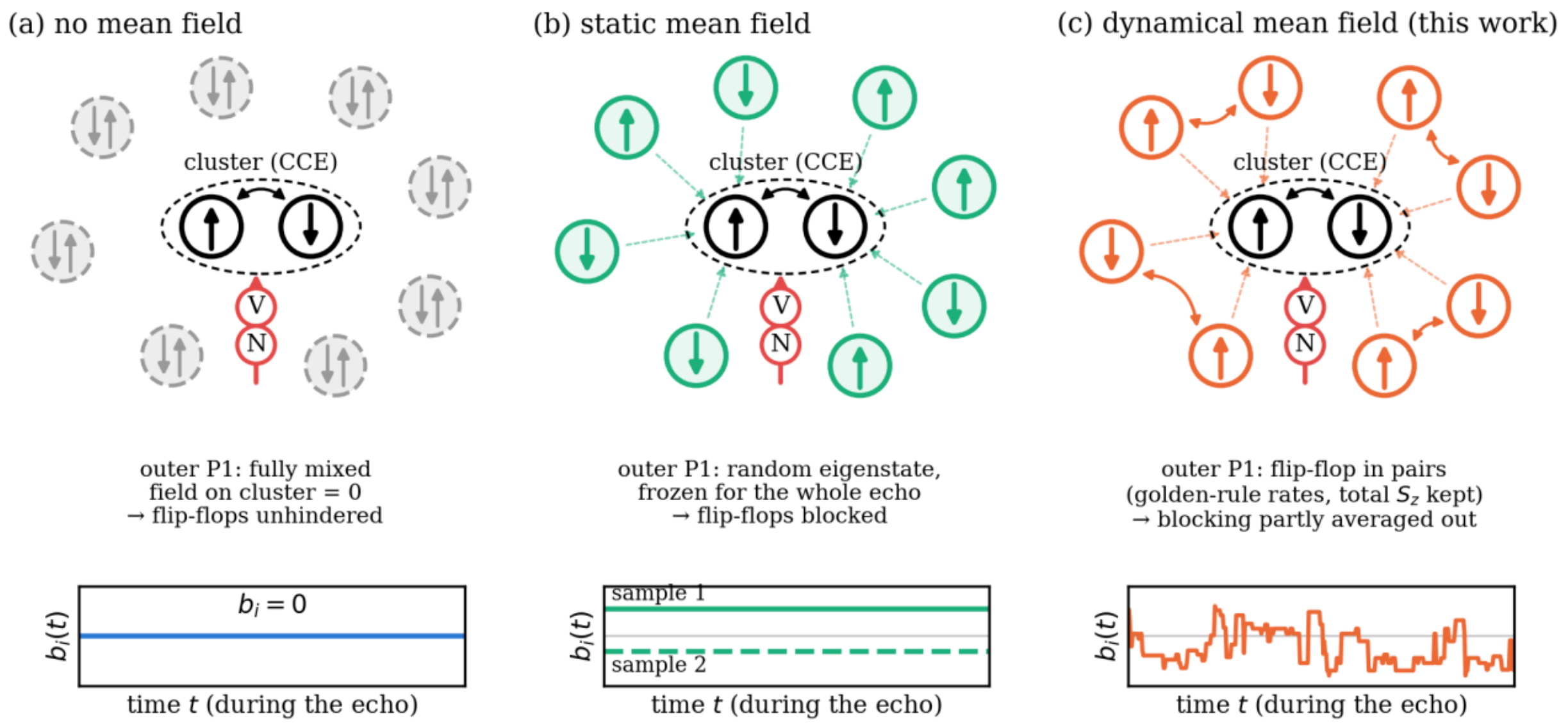

FIG. 1. Three treatments of the P1 centres outside a CCE cluster. A cluster (here a pair of P1 centres, black) flip-flops, and this dephases the NV centre. The outer centres act on the cluster through the Overhauser field $b_i(t)$ of Eq. (2), sketched in the lower panels. (a) No mean field: the outer centres are fully mixed and exert no field, so the pair flip-flops freely. (b) Static mean field: each outer centre is put in a random eigenstate, which stays frozen during the echo. The resulting constant field detunes the pair and blocks its flip-flop; the echo is averaged over samples. (c) Dynamical mean field (this work): the outer centres flip-flop in pairs (curved arrows) and conserve the total $S_z$ of each pair. Their rates are golden-rule rates at the instantaneous detuning. The field therefore fluctuates on the time scale of the echo, and its blocking effect is partly averaged out. Zero rate recovers (b), and infinite rate recovers (a).

## II. MODEL AND METHOD

### A. P1 bath with explicit nitrogen

We treat each P1 centre k as an electron spin S = 1/2 coupled to its own $^{14}$N nucleus (I = 1) [11,18,19]:

$$H_k = -\gamma_e B S_z - \gamma_N B I_z + \mathbf{S} \cdot \mathsf{A}_k \cdot \mathbf{I} + P(\mathbf{n}_k \cdot \mathbf{I})^2 \tag{1}$$

with $\mathsf{A}_k = A_\perp \mathbb{1} + (A_\parallel - A_\perp)\, n_k n_k^T$, $A_\parallel$ = 114.03 MHz, $A_\perp$ = 81.31 MHz and P = −3.97 MHz. The Jahn-Teller axis $n_k$, the axis along which the centre distorts, is drawn at random from the four ⟨111⟩ directions. The field B is parallel to the NV axis. At 2.5 mT the electron Zeeman splitting (70 MHz) is smaller than the hyperfine interaction. The six eigenstates |a⟩ of $H_k$ are then strongly hybridised, i.e. they mix electron and nuclear states. Their electron moments $s_{k,a} = \langle a|S_z|a\rangle$ have a mean modulus of 0.30 at 2.5 mT, which rises to 0.47 at 10 mT and 0.50 at 30 mT. The P1 centres interact with each other and with the NV centre through the point-dipole interaction $J_{kj} = (\mu_0\gamma_e^2\hbar/4\pi r^3)(1 - 3\hat{r}\hat{r}^T)$. The qubit levels of the NV centre are $m_S$ = 0 and −1, and the NV centre enters each bath Hamiltonian through the conditional term $m_S\, J_{NV,k}\, \hat{z}\cdot S_k$. We draw the positions at random at a density of 1 ppm (mean spacing a = 178 Å) inside a sphere of radius 5.5a around the NV centre. Since $T_2 \propto 1/[N]$ in this bath [2], we quote $T_2\cdot[N]$. We verified this scaling for all three treatments: between 0.3 and 10 ppm, the dynamical $T_2\cdot[N]$ stays within 2 % of its 1 ppm value at 2.5, 10, 30 and 50 mT (Appendix, Sec. S6).

### B. Cluster expansion and the Overhauser mean field

We evaluate the Hahn-echo coherence in the CCE to second order (CCE2), with whole P1 centres (electron plus nucleus) as the units. The expansion thus contains single centres and all pairs closer than 4a, and this pair cutoff is converged (Appendix, Sec. S5 and Table S2). For a cluster C, the outer P1 centres act through the Overhauser term

$$H_{\mathrm{MF}}(t) = \sum_{i\in C} b_i\,(t) S_{z,i},\, b_i(t) = \sum_{k\notin C} J_{ik}^{zz}\, s_k(t) \tag{2}$$

In the mean-field-free treatment $s_k = 0$; the outer spins are fully mixed. In the static-mean-field treatment [6], each outer centre is put in one of its eigenstates at random, and $s_k$ stays frozen at the electron moment of that state for the whole evolution. The cluster itself stays fully mixed, so the calculation averages exactly over its entangled internal states.

We stop the expansion at second order for four reasons. First, the elementary process that decoheres the NV centre in this bath is the flip-flop of a P1 pair, and CCE2 treats every such pair exactly. Second, the main effect of a third centre on a pair is its Ising (zz) field, which shifts the pair away from resonance. The Overhauser term of Eq. (2) already contains this field from all outer centres, so much of what CCE3 would add enters through the mean field. The dynamical mean field of Sec. II C goes one step further. It also contains the change of this field when the outer centres flip-flop among themselves; in the cluster expansion, this effect would appear only at third and higher order. Third, the conventional expansion beyond second order is not reliable in strongly coupled electron-spin baths. There it often gives unphysical results such as $|L| > 1$ [7,8], because a cluster contribution is divided by subcluster contributions that pass close to zero. In our P1 bath CCE3 diverges, so a higher-order calculation would not be more accurate in practice. Schätzle et al. tested the convergence with a partitioned CCE, which treats strongly coupled groups of spins as single units [8]. For P1 baths with the nitrogen hyperfine interaction, their expansion converged already at the level that corresponds to CCE2. They attributed this to the hyperfine substructure of the bath, which suppresses flip-flops between centres of different hyperfine subgroups [8]. Fourth, the CCE2 results pass the available checks. The pair cutoff is converged, $T_2$ scales as 1/[N] (Sec. II A), and the static-mean-field limit reproduces the result of Park et al. at 50 mT (Sec. III B). It is also close to the static-mean-field result of Schätzle et al. for bulk diamond, 253 ± 13 µs·ppm [8]; we obtain 264 ± 16 µs·ppm at 30 mT. Correlated flips of three or more centres remain outside CCE2; we return to them in Sec. IV. In the next subsection we let $s_k(t)$ change in time.

### C. Dynamical mean field

The outer P1 centres start from uniformly drawn eigenstates and evolve by flip-flops. In a flip-flop, centre k goes from eigenstate a to a′ and a neighbour j goes from b to b′ at the same time. The matrix element of this transition is $V = \Sigma_{\mu\nu} J^{\mu\nu}_{kj} \langle a'|S_\mu|a\rangle\langle b'|S_\nu|b\rangle$, and its intrinsic energy mismatch is $\Delta^0 = (E_{a'} - E_a) + (E_{b'} - E_b)$. In a bath of other spins, the mismatch also contains the change of the Overhauser energy of the two centres. For the current states of all centres the rate is therefore

$$W_{kj}(ab \to a'b') = \lambda 2\pi |V|^2 g_\sigma\big(\Delta^0 + h_k \delta s_k + h_j \delta s_j + J_{kj}^{zz} \delta s_k \delta s_j\big) \tag{3}$$

where $\delta s_k = s_{k,a'} - s_{k,a}$, $h_k = \Sigma_m J^{zz}_{km}\, s_m(t)$ is the instantaneous field on centre k from all other centres, and $g_\sigma$ is a normalised Gaussian. The width $\sigma = 2|V|$ covers strongly coupled close pairs, for which the golden rule does not apply: they flip at a rate of order $|V|$ rather than $|V|^2/\sigma$. For all other pairs the Gaussian is narrow compared with the detuning, so a pair flip-flops only while the surrounding field holds it near resonance. We include pairs closer than 4a. We exclude transitions with $|\Delta^0| > 10$ MHz; these are non-secular double flips, which do not conserve the Zeeman energy and arise only for the closest pairs.

Kinetic Monte Carlo, a stochastic simulation that draws events one at a time according to their rates, generates the trajectories. The rate without the Gaussian factor is an upper bound on W. This bound depends only on the states of the two partners, and it changes only for pairs that share a centre with the last event. The algorithm therefore draws candidate events from these bounds and accepts each with probability $\exp(-\Delta^2/2\sigma^2)$; this procedure is exact. The rates obey detailed balance at infinite temperature, i.e. forward and backward rates are equal. The uniform distribution of eigenstates therefore stays stationary, as in the static treatment. The factor λ scales all rates, and λ = 1 is the physical, parameter-free choice. For λ = 0 the field is frozen, and the static mean field is recovered exactly. For $\lambda \to \infty$, $s_k(t)$ averages to the eigenstate mean $\mathrm{Tr}\, S_z / d = 0$ within any finite time, and the mean-field-free result is recovered. This is motional narrowing: fast fluctuations average out [16,17].

For comparison we also use a cheaper variant, in which every outer centre jumps on its own. Its rate is the rate of Eq. (3), summed over partners and averaged over the initial state of the partner:

$$W_k(a \to a') = \lambda \sum_j \frac{1}{d} \sum_{b \neq b'} 2\pi\, |V|^2 g_{\sigma'}(\Delta^0) \tag{4}$$

where d = 6 is the number of eigenstates. Here the static spread of the outer field replaces the instantaneous detuning: $\sigma'^2 = \Sigma_m \langle s_m^2 \rangle\, [J^{zz}_{km}\, \delta s_k + J^{zz}_{jm}\, \delta s_j]^2 + (2|V|)^2$. For each centre, these independent jumps form a continuous-time Markov process (a random process without memory) with the same limits. They differ from the pair flip-flops in two respects. First, the two partners of a flip-flop do not change together, so the total $S_z$ is not conserved. Second, the rates do not depend on the instantaneous detuning. The first difference matters more for the field at a cluster. A flip-flop of a distant pair changes this field by $J^{zz}_{ik}\, \delta s_k + J^{zz}_{ij}\, \delta s_j$ with $\delta s_j \approx -\delta s_k$, the difference of two nearly equal dipolar fields. Independent jumps, in contrast, add the two changes at random times. Unless stated otherwise, 'dynamical mean field' refers to the pair flip-flops.

For a given trajectory r of all outer centres, each cluster evolves under the piecewise-constant Hamiltonian $H_C + H_{MF}^r(t)$, and the field is averaged exactly over each time step. The echo coherence is the product $L^r(t) = \Pi_C\, \tilde{L}_C^r(t)$ of the irreducible cluster contributions. The measured coherence is the trajectory average $L(t) = \langle L^r(t) \rangle_r$, and $T_2$ is the 1/e time of its configuration average. We average the product, not the individual factors, so the correlations between clusters that share outer spins are kept. We read $T_2$ directly from the decay, without a fit. A stretched-exponential fit, $|L| = \exp[-(2\tau/T_2)^p]$, gives the same $T_2$ within 2 % for the pair flip-flops and within 4 % for all treatments (Appendix, Sec. S8 and Table S4).

### D. Implementation and validation

All calculations in this work, including the two limits, were done with an original NumPy code that we developed for this work. It implements the cluster-correlation expansion with a dynamical mean field, and we have packaged it as dynacce. The package reproduces the results of the production code to $\leq 5\times10^{-11}$ in the two limits and to $\leq 1\times10^{-9}$ with the dynamical mean field (Sec. S1). The code is not publicly released and is available from the author upon request. In the two limits our code reproduces PyCCE 1.1.1 [12] on the same configurations. Over all configurations and fields, |L| agrees to $7\times10^{-11}$ without mean field and to $\leq 1.2\times10^{-3}$ with the static mean field, and $T_2$ agrees to $2\times10^{-5}$ (Appendix, Sec. S1 and Fig. S1). The numerical techniques that make the dynamical calculation affordable are described in Secs. S2–S4. We use 8 trajectories per configuration and 23 of 24 random bath configurations. We exclude one configuration, in which a P1 centre sits 0.08a from the NV centre and CCE2 does not converge (Sec. S5).

The errors are standard deviations from a bootstrap over configurations (a resampling of the configurations), paired between treatments and fields.

## III. RESULTS

### A. How fast does the Overhauser field fluctuate?

Figure 2(a) shows the distribution of the total flip-flop rate of each centre, $\Gamma_k = \Sigma_{a'} W_k(a \rightarrow a')$ of Eq. (4), averaged over the states a, for P1 centres in the interior of the bath. The distribution is broad and spans more than three decades. The median is 14 $ms^{-1}$ at 2.5 mT and 4.7 $ms^{-1}$ at 30 mT. At 2.5 mT the central 80 % lie between 1.7 and 94 $ms^{-1}$, and a tail of strongly coupled close pairs flips much faster. The rates are higher at low field because of hybridisation. It gives $S_x$ and $S_y$ matrix elements between many more pairs of eigenstates. These matrix elements open resonant channels between centres of different Jahn-Teller orientation, and these channels are closed at high field.

What matters for the pairs, however, is the Overhauser field $b_i(t)$ itself. This field is a sum over many centres weighted by $1/r^3$, and its rms amplitude is 266–409 kHz at 1 ppm. Figure 2(b) shows its autocorrelation, which measures how long the field remembers its value. We normalise it for each centre i and then average, so the few centres with a close partner, whose field reaches the MHz scale, do not dominate. Under pair flip-flops the field loses its memory gradually, without a fast initial drop. 87–93 % of the correlation remains after 20 μs, and 68–83 % after 200 μs, about the echo time at $T_2$. The correlation falls to 1/e only on the millisecond scale. During the echo the field is therefore largely, but not completely, frozen. The part that changes between the two halves of the echo is enough to dephase the pair signal. The decay is fastest at 2.5 mT, where the rates in Fig. 2(a) are highest. When every centre jumps on its own, the correlation falls to 1/e after 102–106 μs, and only 28–28 % is left at 200 μs. Independent partners thus make the field lose its memory more than ten times faster, because the changes of $b_i$ from the two partners no longer partly cancel. This is the origin of the shorter $T_2$ of the independent-jump approximation in Sec. III B. The real Overhauser field is thus neither frozen nor motionally narrowed (averaged out by fast fluctuations). Its correlation time is comparable to or longer than $T_2$. In this regime its dynamics must be computed rather than assumed. This anticipates the result of the next section: the dynamical $T_2$ lies between the two limits, closer to the static one.

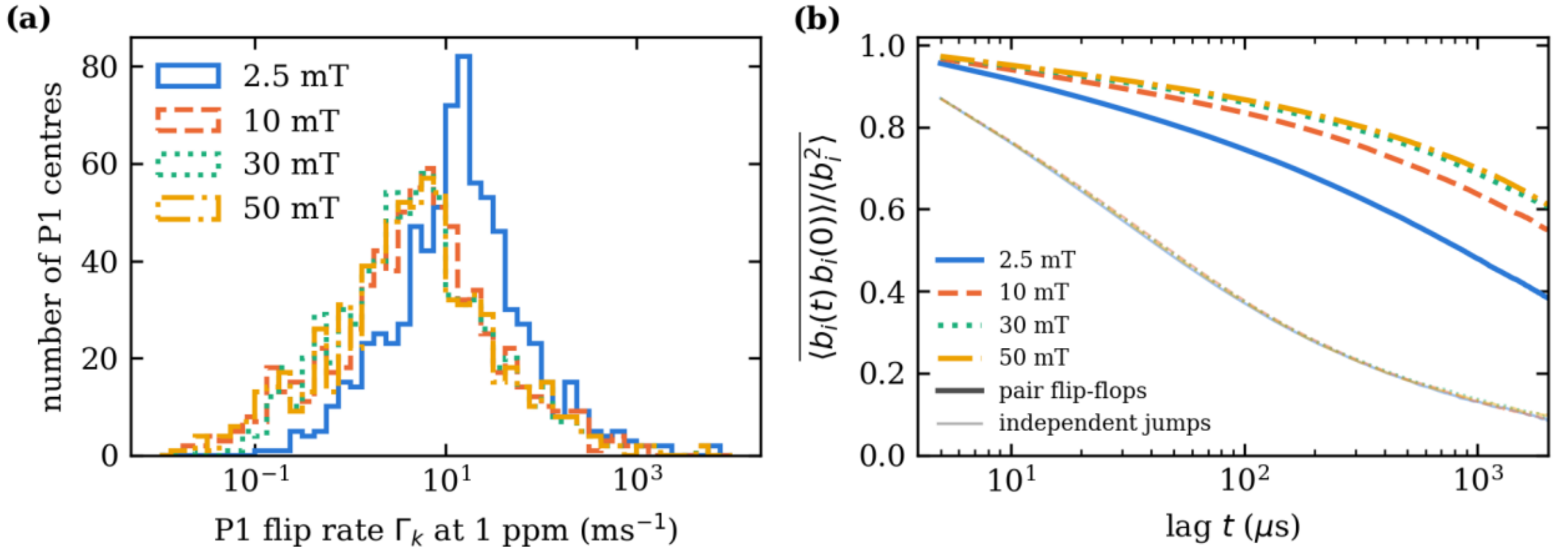

FIG. 2. Dynamics of the outer P1 bath at 1 ppm ($^{14}N$, λ = 1). (a) Distribution of the total golden-rule jump rate $\Gamma_k$ of P1 centres within 3a of the NV centre (seven configurations per field). (b) Autocorrelation of the Overhauser field $b_i(t)$ on those centres, normalised for each centre and averaged, on a logarithmic time axis. Thick lines show pair flip-flops (16 trajectories per configuration), and thin lines show independent jumps; the thin curves for the four fields nearly coincide.

### B. Echo decay and comparison with 160 μs·ppm

Figure 3 compares the configuration-averaged echo decays of the treatments. At 30 mT [Fig. 3(b)], the static mean field gives $T_2 \cdot [N] = 264 \pm 16$ μs·ppm. This is 1.6 times the measured $160 \pm 12$ μs·ppm, in line with the factor of two that Park et al. found [6]. Without mean field, $T_2 \cdot [N] = 68 \pm 3$ μs·ppm, a factor 2.4 too short. The dynamical mean field gives $188 \pm 12$ μs·ppm, 1.7 combined standard deviations above the measured value. We call the fraction of the gap between the two limits that remains the position (0 = no mean field, 1 = static). The fluctuating field gives a position of $0.62 \pm 0.03$. It thus undoes about 38 % of the static enhancement and reduces the excess over experiment from a factor 1.6 to 1.18. Within the P1 bath alone, and without adjustable parameters, most of the factor-of-two excess at 30 mT therefore disappears. At 10 mT the dynamical result, $167 \pm 11$ μs·ppm, agrees with experiment ($+0.4\sigma$).

The independent jumps of Eq. (4) give shorter echo times: $147 \pm 9$ μs·ppm at 30 mT ($0.9\sigma$ below experiment). On the same configurations and initial states, the pair flip-flops lengthen $T_2$ by a factor $1.28 \pm 0.03$ at 30 mT, $1.20 \pm 0.03$ at 10 mT and only $1.09 \pm 0.01$ at 2.5 mT. Independent jumps thus over-randomise the field. They let each partner of a flip-flop change the field at a cluster on its own. In a real flip-flop, however, the two changes are simultaneous and opposite, and they partly cancel. The field at the cluster is therefore more persistent than independent jumps make it. The static mean field freezes the field too much, and the independent jumps freeze it too little. The pair flip-flops keep the $S_z$ conservation of the underlying dynamics and lie between them. Independent jumps remain a useful approximation. They cost less, they are sufficient where hybridisation keeps the field weak (2.5 mT), and their $T_2$ errs on the short side at every field. The 20–30 % that they miss at high field, however, is comparable to the remaining difference from experiment. For comparison, the Hahn-echo times that Park et al. measured at 28–30 mT correspond to 131 and 137 μs·ppm for their 0.8 and 13 ppm samples [6], somewhat below the benchmark of Ref. [2].

Park et al. report a static-mean-field $T_2 \cdot [N]$ of 286 μs·ppm at 50 mT, computed with explicit nitrogen [6]. At the same field our static mean field gives $266 \pm 16$ μs·ppm, within 1.2 standard deviations of their value, so the two implementations agree in this limit. The dynamical mean field gives $190 \pm 11$ μs·ppm, 0.71 times the static value. The fluctuating field thus removes the same fraction of the static enhancement at the field where the overestimate was first reported.

At 2.5 mT [Fig. 3(a)] the situation differs. The treatments lie closer together, because hybridisation reduces the electron moments that generate the field. The dynamical result, $112 \pm 8$ μs·ppm, lies $3.3\sigma$ below the measured value; its position is $0.76 \pm 0.05$. Even the static limit, $123 \pm 10$ μs·ppm, stays below 160 μs·ppm, and this limit bounds every dynamical result from above. At low field the P1-only calculation therefore underestimates $T_2$, whatever the rate of the field fluctuations. Table I collects the values.

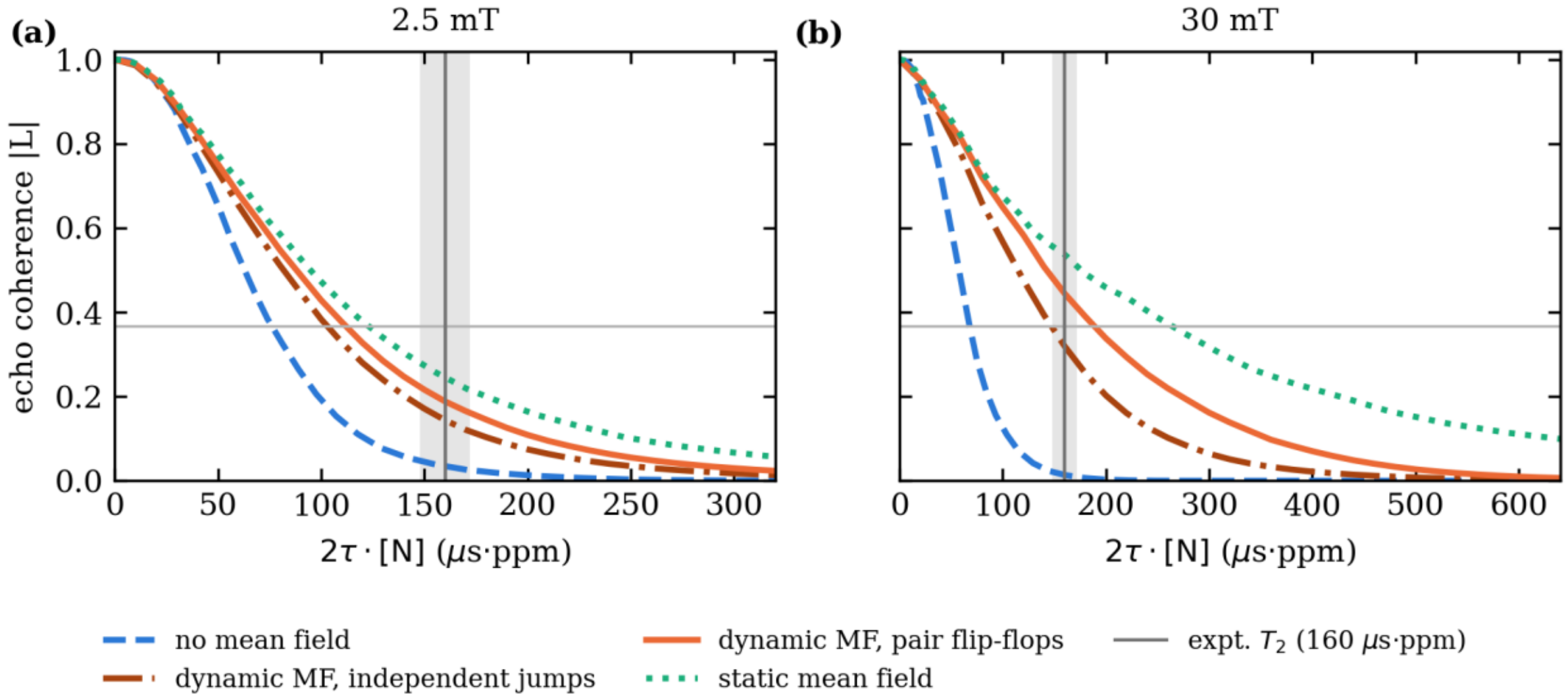

FIG. 3. Configuration-averaged Hahn-echo decays of an NV− centre in a $^{14}N$ P1 bath, as a function of the total echo time multiplied by the concentration, at (a) 2.5 mT and (b) 30 mT. The curves show CCE without mean field (dashed), the dynamical mean field at the golden-rule rates with pair flip-flops (solid) and with independent jumps (dash-dotted), and the static mean field (dotted). The vertical band marks $T_2$·[N] = 160 ± 12 μs·ppm of Ref. [2], and the grey horizontal line marks 1/e.

TABLE I. $T_2$·[N] (μs·ppm) of an NV− centre in a $^{14}N$ P1 bath for each treatment of the Overhauser field (23 configurations; λ = 1 for the dynamical ones). 'Position' is $(T_{pair} - T_{none})/(T_{static} - T_{none})$. The last column compares the pair-flip-flop result with 160 ± 12 μs·ppm [2].

| B (mT) | no mean field | independent jumps | pair flip-flops | static | position | pair / 160 |
|---|---|---|---|---|---|---|
| 2.5 | 76 ± 5 | 103 ± 8 | 112 ± 8 | 123 ± 10 | 0.76 ± 0.05 | 0.70 (−3.3σ) |
| 10 | 70 ± 3 | 138 ± 9 | 167 ± 11 | 228 ± 18 | 0.61 ± 0.03 | 1.04 (+0.4σ) |
| 30 | 68 ± 3 | 147 ± 9 | 188 ± 12 | 264 ± 16 | 0.62 ± 0.03 | 1.18 (+1.7σ) |
| 50 | 68 ± 3 | 146 ± 9 | 190 ± 11 | 266 ± 16 | 0.62 ± 0.03 | 1.19 (+1.8σ) |

## C. Field dependence

Figure 4 summarises the field dependence. From 2.5 to 30 mT the mean-field-free $T_2$ stays flat, the static one grows by a factor 2.15 ± 0.08, and the dynamical one grows by 1.69 ± 0.07 (167 ± 11 μs·ppm at 10 mT and 188 ± 12 μs·ppm at 30 mT and 190 ± 11 μs·ppm at 50 mT). Bauch et al., in contrast, measured a constant $T_2$ within about 10 % over 2–30 mT [2]. The dynamical field reduces $T_2$ most where the static enhancement is largest: by 29 % of the static value at 30 mT, but by 9 % at 2.5 mT. Part of the spurious field dependence of the static treatment therefore disappears. Independent jumps give a weaker rise, 1.43 ± 0.04, because the pair correlation that they neglect matters most at high field (Sec. III B). Their flatter field dependence does not indicate a better description; it only reflects a smaller $T_2$ at high field. The remaining field dependence lies between 2.5 and 10 mT. Above 10 mT the dynamical $T_2$ is nearly constant. It follows the electron moment of the P1 eigenstates, which saturates (0.47 at 10 mT, 0.50 at 30 mT). At 2.5 mT hybridisation reduces this moment to 0.30, and with it the Overhauser field that suppresses the pair flip-flops. The measured $T_2$ shows no such step. What the P1-only description still misses therefore lies in the regime of strong electron–nuclear hybridisation.

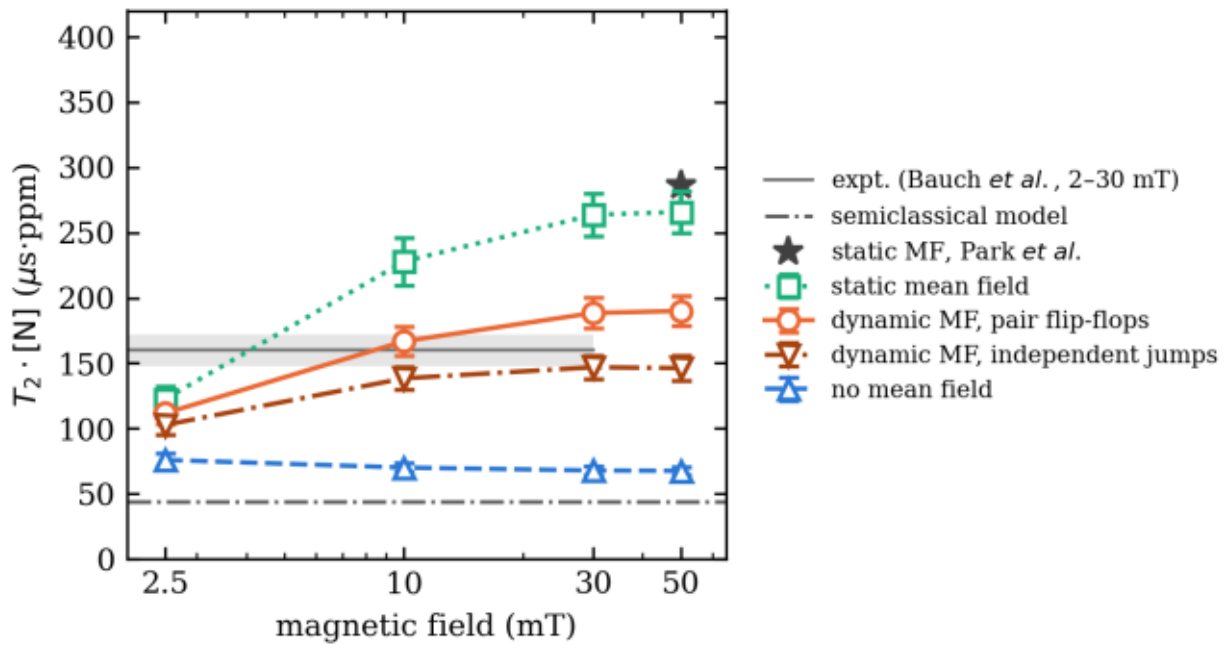

FIG. 4. $T_2$·[N] versus bias field for the treatments of Fig. 3 ($^{14}N$, 23 configurations). The grey band is the field-independent 160 ± 12 μs·ppm measured over 2–30 mT [2]; the star is the static-mean-field result of Park et al. at 50 mT [6]; the dash-dotted line is the semiclassical model, 44 μs·ppm [3].

## D. Sensitivity to the jump rates

The golden-rule rates rest on the approximations of Sec. II C. We therefore scaled them by $\lambda = 0.3$ and 3 for 23 of the configurations (Fig. 5). For this sweep we used the cheaper independent jumps of Eq. (4), which share the matrix elements and line shapes of the pair flip-flops. $T_2$ decreases monotonically with $\lambda$ between the two limits, but only weakly. Relative to $\lambda = 1$ on the same configurations, $T_2$ changes by +14 ± 2 % ($\lambda = 0.3$) and −11 ± 1 % ($\lambda = 3$) at 30 mT, and by +2 ± 1 % and −4 ± 1 % at 2.5 mT. A factor-of-three uncertainty in the rates therefore moves $T_2$ by about 10 %. This change is smaller than the effect of the partner correlation (Sec. III B) and an order of magnitude smaller than the static enhancement. We then repeated the test with the pair flip-flops themselves, for seven configurations (Appendix, Sec. S9 and Table S5). Besides $\lambda = 0.3$ and 3, we halved and doubled the width $\sigma$ of the line shape in Eq. (3). At 30 mT, $T_2$ changes by between −10 % and +6 %. At 2.5 mT, every change is within 1 %, inside the statistical error. With these factors, the 23-configuration results span 169–200 μs·ppm at 30 mT and 110–112 μs·ppm at 2.5 mT. The first range lies 6–25 % above the measured central value, and its lower end falls within the experimental uncertainty, 160 ± 12 μs·ppm [2]. The second range stays below 160 μs·ppm. The low-field shortfall therefore depends neither on the precise rates nor on the line width.

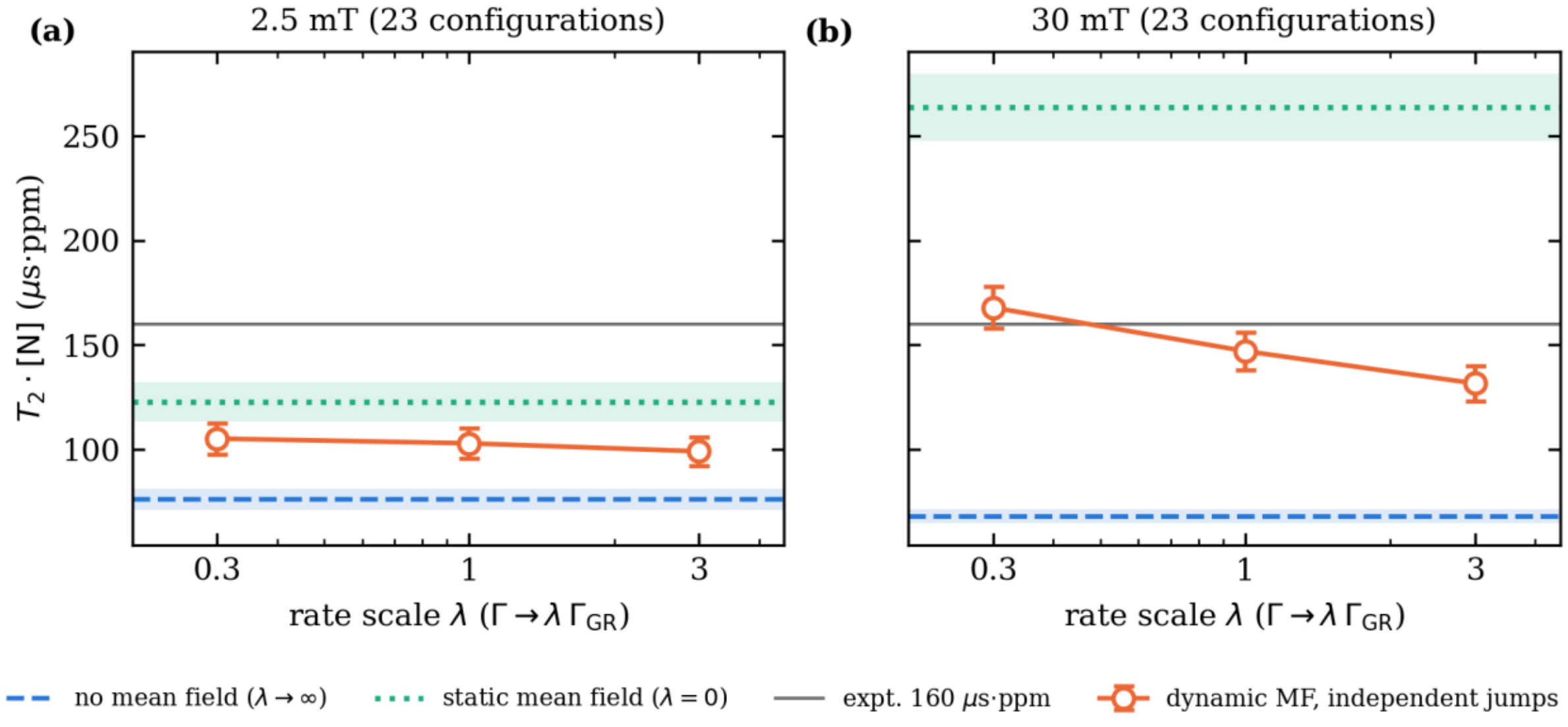

FIG. 5. $T_2$·[N] versus the rate scale λ (independent jumps, Eq. (4)) at (a) 2.5 mT and (b) 30 mT, for the 23 configurations of the sweep. Horizontal lines and bands show the no-mean-field result ($\lambda \to \infty$, dashed) and the static mean field ($\lambda = 0$, dotted) for the same configurations; the solid grey line marks 160 µs·ppm [2]. Error bars show the spread between configurations. The changes with λ on the same configurations are much better determined (see text).

## IV. DISCUSSION

The central result is that static-mean-field CCE overestimates $T_2$ in a P1 bath mostly because it freezes the Overhauser field. The frozen picture is not wrong in itself. Under pair flip-flops, 68–83 % of the field correlation survives the echo time [Fig. 2(b)], so the field is largely quasi-static. The small part that does change between the two halves of the echo, however, lowers $T_2$·[N] at 30 mT from 264 to 188 µs·ppm. It thus removes about 73 % of the excess over experiment. Park et al. [6] matched their Hahn-echo $T_2$, measured at 28–30 mT and computed at 50 mT, with dark paramagnetic defects at fitted concentrations. We remove most of the same excess without additional defects and without fitting: we only let the P1 bath that produces the field evolve. The two explanations do not exclude each other. Dark spins certainly exist in CVD diamond and decohere NV centres [10], and in samples where they are abundant they will shorten $T_2$ further. Our result means, however, that the excess of a static-mean-field calculation over experiment cannot measure their concentration, because most of that excess is not real. Moreover, as noted in Sec. I, the NV centres of those NV-ensemble samples already decohere the ensemble by about as much as the fitted dark spins [10], and a P1-only model omits them. Part of what has been attributed to unknown defects may therefore be the NV ensemble itself. A quantitative treatment requires the orientation distribution, charge states and driving conditions of the experiment, and it is beyond the present work.

Together with the semiclassical model, the four treatments form a hierarchy: they differ in how long the local field of the surrounding bath persists (Fig. 4). If the field acts as fast homogeneous dephasing, $T_2$·[N] is 44 µs·ppm [3]. Without the field it is 68 µs·ppm. With independent jumps it is 147 µs·ppm, with correlated pair flip-flops 188 µs·ppm, and with a frozen field 264 µs·ppm at 30 mT. The measured value is 160 µs·ppm. $T_2$ rises monotonically along this sequence, because a more persistent field detunes more pair flip-flops for longer. The measured value lies within the range of the two computed field dynamics, far from both limits. $T_2$ therefore measures, more than anything, how quasi-static the Overhauser field is. This explains why the semiclassical model already reproduces $T_2^*$, which is insensitive to this question [3]. The step from independent jumps to pair flip-flops shows that the persistence of the field depends not only on how often the bath spins flip but also on how they flip. When each flip-flop conserves $S_z$, the fields of the two partners partly cancel at a distant cluster, and the field becomes correspondingly more static.

At high field the pair flip-flops overshoot the benchmark moderately: by 18 % at 30 mT (1.7σ) and by 19 % at 50 mT. An excess of this size leaves room for other nitrogen-related paramagnetic centres in real samples, which a P1-only model omits. $NV^-$ and $NV^0$ centres are examples. Their concentration in nitrogen-rich diamond grows with [N], and they decohere the ensemble more strongly per centre than P1 [10]. Because their concentration scales with [N], the empirical constant would absorb them, and they would not appear as a sample-dependent excess. The excess is also within the range of the remaining approximations of the bath dynamics, which we discuss below. What the calculation does exclude is a factor of two: once the bath can move, the dark spins that repair a static-mean-field calculation are not needed.

At low field the calculation errs in the opposite direction. Even the frozen field leaves $T_2$ below 160 μs·ppm at 2.5 mT, so further decoherence sources cannot repair the result; what is needed is less decoherence or a different reference. Several approximations act precisely where hybridisation is strong. (i) The golden-rule rates treat the multilevel electron–nuclear eigenstates as independent channels. They use a Gaussian line shape whose width the flip-flop coupling alone sets. At 2.5 mT many channels of similar energy overlap, and both the Markov assumption (no memory between events) and the rate magnitudes are least secure there. A factor of three in the rates or a factor of two in the line width, however, moves $T_2$ by less than 1 % at 2.5 mT (Sec. III D). The rate magnitudes are therefore not the cause; the treatment of overlapping channels as independent remains untested. (ii) The bath dynamics consists of pair flip-flops only. Correlated flips of three or more centres connect more eigenstates at low field, but they are absent both from the trajectories and from CCE2. (iii) On the experimental side, the concentrations of Ref. [2] are total nitrogen, whereas the bath consists of neutral substitutional nitrogen. If only a fraction f of the nitrogen is P1, the benchmark per P1 centre is 160f μs·ppm. This would shrink the low-field discrepancy and enlarge the high-field overshoot, but it cannot remove the field dependence. The partner correlation, which lengthens $T_2$ by 20–30 % at high field, does little here (a factor 1.09 at 2.5 mT). The low-field shortfall therefore does not come from the way we generate the trajectories. The low-field regime calls for both a better description of the hybridised bath and a field-resolved measurement on samples with independently determined P1 concentration.

The method itself is general. It requires only the eigenstates of the bath units and their pairwise couplings, and its limits are the two standard mean-field treatments. It uses a single physical parameter set and gives the correlation time of the Overhauser field explicitly. It should therefore help wherever a dilute electron bath decoheres a central spin on the time scale of its own flip-flops. Dynamical-decoupling sequences are one example: their filter function, the frequency window that a pulse sequence probes, measures that correlation time directly [13,14].

## V. CONCLUSIONS

We have introduced a dynamical mean field into the cluster-correlation expansion of NV− decoherence in a P1 spin bath. In this model the outer P1 centres flip-flop in pairs, conserve their total $S_z$ and follow golden-rule rates without adjustable parameters. The dynamical mean field reduces the static-mean-field $T_2$·[N] at 30 mT from 264 to 188 µs·ppm and gives 167 µs·ppm at 10 mT, against the benchmark 160 ± 12 µs·ppm. The factor-of-two excess of static-mean-field calculations is therefore mostly an artefact of the frozen Overhauser field, not evidence for additional defects. When the centres jump independently, a cheaper approximation, $T_2$·[N] is 147 µs·ppm at 30 mT; the correlation of the flip-flop partners lengthens $T_2$ by 20–30 % at high field. At 2.5 mT the calculation gives 112 µs·ppm, and even its static limit stays below 160 µs·ppm. A residual increase of $T_2$ with field, a factor 1.69 between 2.5 and 30 mT, therefore remains where experiment finds none. The fluctuating Overhauser field explains most, but not all, of the gap between CCE and experiment.

More generally, the treatment of the spins outside each cluster is not a technical detail of the CCE; it is part of the physical model of the bath. A static mean field assumes that these spins are frozen, and the mean-field-free treatment assumes that they are fully randomised. A bath of flip-flopping electron spins lies in between, and its own dynamics sets where. The dynamical mean field derives that dynamics from the bath Hamiltonian alone: which centres can flip-flop, with which partners and at which rates. It then couples this dynamics to the cluster expansion without adjustable parameters. The method therefore carries over directly to other electron-spin baths and pulse sequences. Examples are denser defect ensembles, other paramagnetic defects in diamond and in other host materials, mixed electron–nuclear baths, and dynamical-decoupling sequences, which probe the spectrum of the outer-field fluctuations more selectively than the Hahn echo. We expect that CCE calculations of spin coherence in electron-spin baths will need to treat the outer spins dynamically rather than in either static limit.

## DATA AND CODE AVAILABILITY

We performed all calculations with an original code that the author developed and packaged as dynacce; it is not publicly released. The code and the data that support the findings of this study are available from the author upon reasonable request.

## APPENDIX: SUPPLEMENTAL MATERIAL

*The sections, figures and tables below are labelled S1–S9, Figs. S1–S6 and Tables S1–S5 as cited in the main text.*

## APPENDIX S1: VALIDATION AGAINST PYCCE

dynacce draws the P1 positions with the same random generator and seed as random_bath of PyCCE [12], so the two codes are compared on identical configurations. The bath contains the Jahn-Teller axes from the same random stream, the hyperfine tensor of each centre as an explicit electron–nucleus interaction, and point-dipole couplings between all other spin pairs, including the electron–nucleus and nucleus–nucleus terms of different centres. The NV centre enters through the conditional projection of its $m_S = 0$ and $-1$ states, and the mean field of the outer spins through the zz component of the dipolar tensor, as in PyCCE. The echo coherence of a cluster is $L_C = \mathrm{Tr}[U_\beta^\dagger\, U_\alpha^\dagger\, U_\beta\, U_\alpha]/d$ with $U_{\alpha,\beta} = \exp(-iH_{\alpha,\beta}\, t/2)$. For configuration 0 at 2.5 mT, Fig. S1, the full CCE2 coherence agrees with PyCCE to $1.1\times10^{-7}$ without mean field and to $\leq 2.2\times10^{-5}$ with the static mean field sampled from the same random stream, over the whole time window. The residual of the static case comes from two conventions of PyCCE: for an odd number of π pulses it returns the complex conjugate of the coherence, and its mean field also acts on the nuclei of the cluster through the electron–nucleus zz coupling. With both conventions adopted the two codes agree to $5\times10^{-12}$; neither changes $|L|$ or $T_2$ by more than $10^{-5}$. We repeated the comparison for all 23 configurations at 2.5, 10 and 30 mT. Without mean field $|L|$ agrees to $7\times10^{-11}$. With the static mean field $|L|$ of a single sample agrees to $\leq 1.2\times10^{-3}$. The electron–nucleus convention accounts for this residual: with it adopted, the sample with the largest residual agrees to $1\times10^{-11}$. The configuration-averaged $T_2$ of the two codes agree to $2\times10^{-5}$. The production calculations used the research code from which the dynacce package was built. In 5 test cases, which include the sample with the largest residual, the two limits from the package agree with the production results to $\leq 5\times10^{-11}$. For one full configuration (2.5 mT, 56 052 pairs) the dynamical mean field with independent jumps agrees to $\leq 1\times10^{-9}$, with the secular pair propagation of Sec. S3. Exact pair propagation has since become the default of the package, so reproducing our results requires selecting the secular option. The pair flip-flop trajectories came from the package itself.

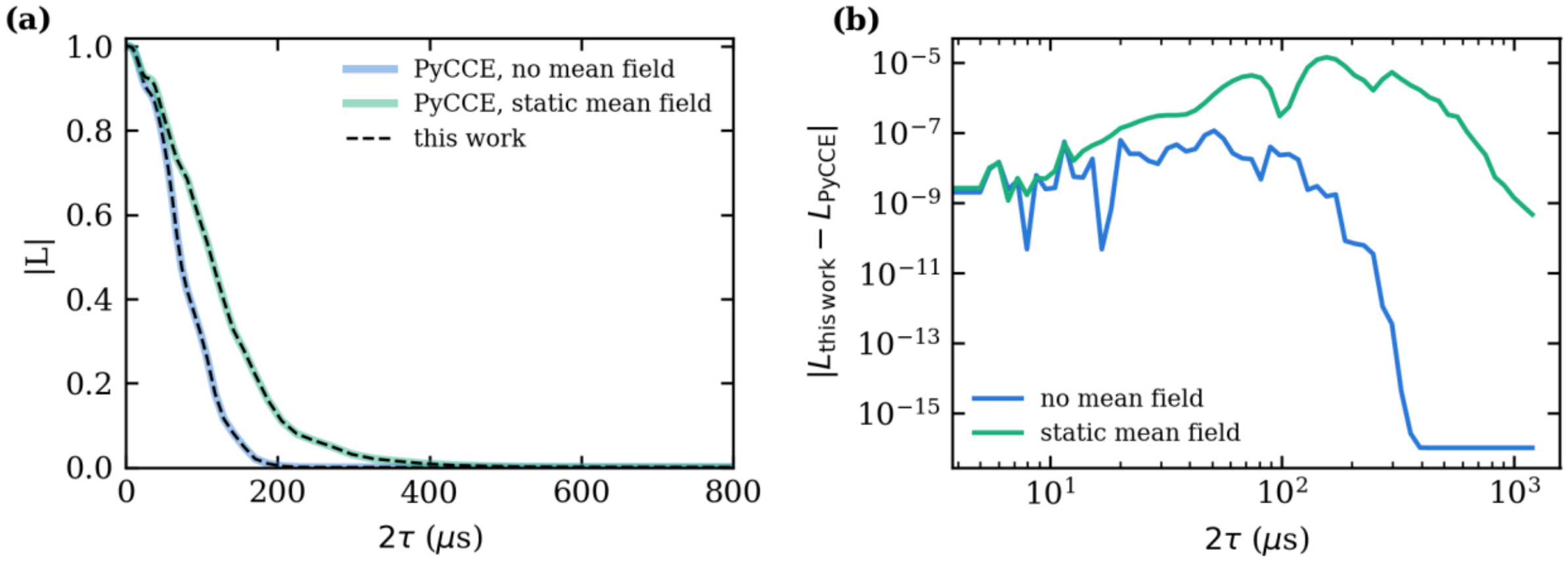

FIG. S1. (a) Echo coherence of configuration 0 at 2.5 mT, $^{14}N$, computed with PyCCE (thick light lines) and with the present implementation (dashed), without mean field and with one static mean-field sample. (b) Absolute difference between the two.

## APPENDIX S2: CLOSED-FORM STATIC ECHO

Three techniques make the dynamical calculation affordable for the ~5×10$^4$ pairs per configuration: a closed-form static echo (this section), secular propagation of the pairs in a time-dependent field (Sec. S3), and a screening that treats only field-sensitive pairs dynamically (Sec. S4). For time-independent conditional Hamiltonians $H_\alpha = W_\alpha E_\alpha W_\alpha†$ and $H_\beta = W_\beta E_\beta W_\beta†$, define the overlap $X = W_\alpha† W_\beta$. With $\tau = t/2$,

$$L(\tau) = \frac{1}{d} \sum_{ik} e^{i\tau(E_{\beta,i} - E_{\beta,k})} |G_{ik}(\tau)|^2, G(\tau) = X^\dagger e^{iE_\alpha\tau} X \tag{S1}$$

G for all time points follows from one matrix product of the $d^2$×d array $X^*_{ji}$ $X_{jk}$ with the d×T array of phases, which is about an order of magnitude faster than forming the propagators. Equation (S1) agrees with direct evaluation to machine precision.

## APPENDIX S3: SECULAR PROPAGATION IN A TIME-DEPENDENT FIELD

In the dynamical treatment a pair Hamiltonian $H_0 + b_1(t)\, S_{z,1} + b_2(t)\, S_{z,2}$ changes at every time step ($\Delta t$ = 5 μs at 2.5 mT and 10 μs at 30 mT; the field of each step is the exact time average of the jump trajectory over that step). Diagonalising it at every step and for every trajectory is the dominant cost. Since |b| (tens of kHz) is small compared with the spacing of most levels of $H_0$ (MHz), we work in the eigenbasis of $H_0$ for each conditional branch, group levels whose spacing is below g = max(0.3 MHz, 30 max|b|) into blocks, keep the field exactly inside the blocks and include its coupling between blocks to second order as a level shift, $\delta E_i = \Sigma_{j\notin blk(i)} |b_1 (S_{z,1})_{ij} + b_2 (S_{z,2})_{ij}|^2/(E_i - E_j)$. Products of step propagators are accumulated in each branch's eigenbasis and combined at the end through X as in Eq. (S1). On pairs sampled from the full range of separations and fields, including strongly coupled close pairs with |b| up to 1.5 MHz, the pair coherence agrees with exact step-by-step propagation to ≤ 5×10$^{-6}$ at three to four times lower cost. The second-order shift is essential: without it the phase error of distant pairs reaches order unity. The small errors of many pairs can add up, so we also tested the approximation on the full calculation. For 12 configurations with pair flip-flops, we repeated the pair terms with exact step-by-step propagation and kept the same bath states, screening and trajectories. The ratio of $T_2$ (exact / secular) is 1.022 ± 0.008 at 2.5 mT and 1.001 ± 0.001 at 30 mT. The $T_2$ with pair flip-flops at 2.5 mT in this work is therefore about 2 % too short. This shift is smaller than the statistical error of that $T_2$ (8 %). All results in the main text use the secular propagation.

## APPENDIX S4: SCREENING OF PAIRS

Most of the ~5×10$^4$ pairs of a configuration are insensitive to the Overhauser field. For each pair C we compute the irreducible contribution $\tilde{L}_C$ without mean field and with the static mean field for 4 of the trajectory initial states, and define its sensitivity $s_C = \max |\tilde{L}_C^{static} / \tilde{L}_C^{none} - 1|$ over these states and over times up to 60 % of the window (which contains the 1/e point). Pairs are treated dynamically in order of decreasing $s_C$ until the sum of the remaining sensitivities falls below 0.02; the remaining pairs enter with their mean-field-free contribution. Figure S2 shows the remaining sum as a function of the number of dynamical pairs. On average 9184 of 56245 pairs are treated dynamically at 2.5 mT and 20987 of 56245 at 30 mT. As an end-to-end test, the screened, step-propagated calculation with $\lambda = 0$ reproduces the unscreened static-mean-field coherence of the same initial states to $|\Delta L| \leq 0.013$ (configuration 0, 2.5 mT), consistent with the tolerance. The kinetic Monte Carlo module of dynacce generates the pair flip-flop

trajectories from the same bath units, eigenstates and couplings. Both kinds of trajectory start from the same initial states and use the same screening, so they differ only in the bath dynamics.

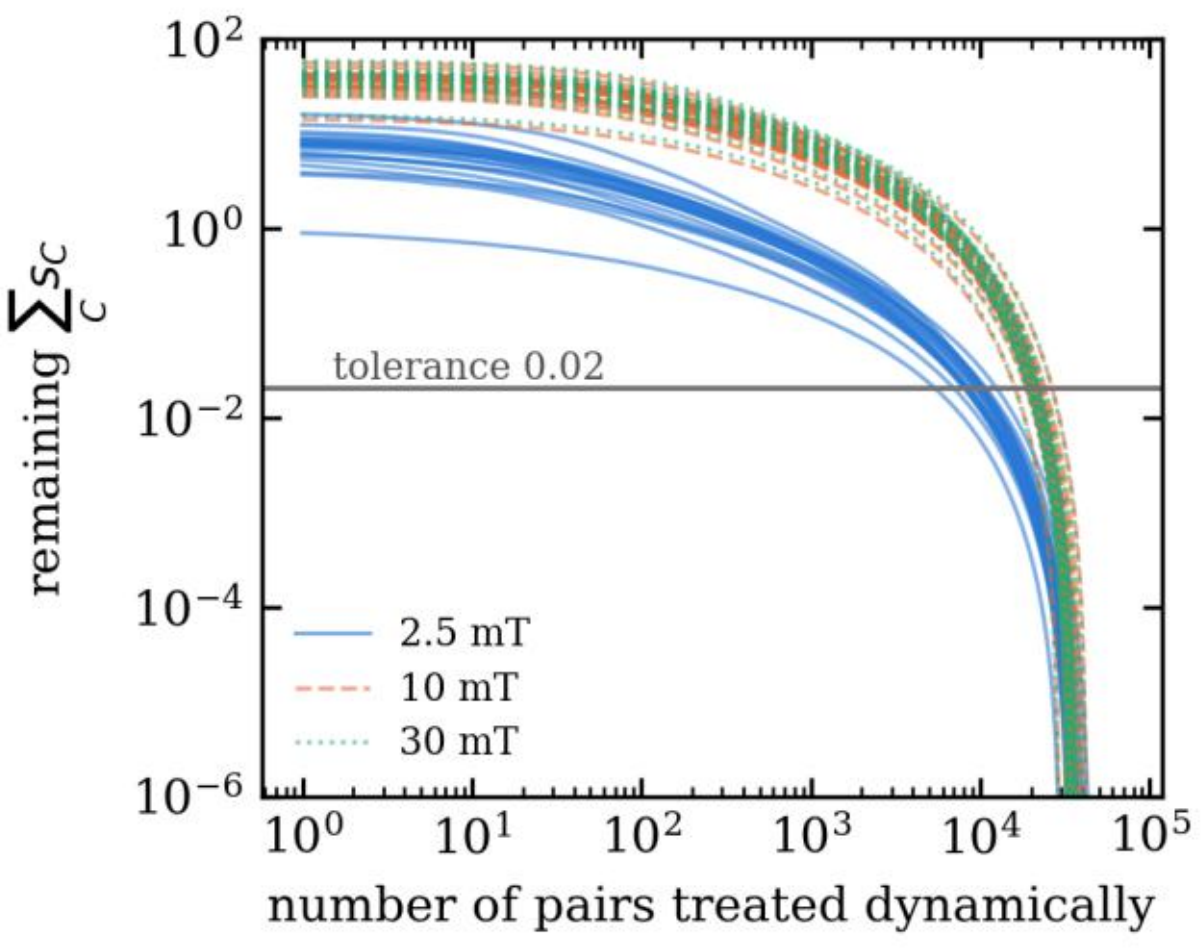

FIG. S2. Sum of the sensitivities of the pairs not treated dynamically, versus the number treated dynamically, for each configuration at 2.5 mT (solid) and 30 mT (dotted). The horizontal line is the tolerance 0.02.

## APPENDIX S5: NUMERICAL PARAMETERS

TABLE S1. Parameters of the calculation (1 ppm).

| **quantity** | **value** |
|---|---|
| P1 hyperfine A∥, A⊥ ($^{14}N$) | 114.03, 81.31 MHz |
| $^{14}N$ quadrupole P | −3.97 MHz |
| mean P1 spacing a | 178 Å |
| bath radius / pair cutoff | 5.5a / 4a |
| CCE order | 2 (in units of whole P1 centres) |
| configurations | 23 of 24, one excluded (paired across treatments and fields) |
| bath dynamics | pair flip-flops (kinetic Monte Carlo, Eq. (3)); independent jumps (Eq. (4)) for comparison |
| flip-flop pair range / secular cut | 4a / $\lvert\Delta^0\rvert \le 10$ MHz |
| trajectories per configuration | 8 |
| time steps | 2K = 64 (Δt = 5 μs at 2.5 mT, 10 μs at 10, 30 mT) |
| screening tolerance | 0.02 |
| code | original code, packaged as dynacce 0.1.0.dev0 (available on request; Sec. S1) |
| $T_2$ estimator | 1/e time of the running minimum of $\lvert\langle L\rangle\rvert$ |
| error bars | bootstrap over configurations (4000 resamples) |

One of the 24 configurations is excluded throughout. In it a P1 centre sits at 0.08a from the NV centre, so that its coupling to the NV spin is comparable to the hyperfine splittings and its single-centre echo alone drops to about 0.5 at 1 ppm. CCE2 then relies on large pair corrections ($|L_C| > 1$ for the pairs containing this centre) to compensate the single-centre factor, and at 3 and 10 ppm, where this coupling grows with [N] while the hyperfine energies stay fixed, the single-centre echo nearly vanishes and the pair corrections diverge. Capping the pair corrections at $|L_C| \leq 1$ is not a remedy: it removes the legitimate compensation and halves $T_2$ of this configuration already at 1 ppm. A P1 centre this close occurs for about 0.2 % of NV centres, so it is over-represented by an order of magnitude in a sample of 24; at 1 ppm removing it changes the $T_2$ values of Table I by less than 1 %.

The pair cutoff 4a was chosen from a convergence test at 1 ppm, 2.5 and 30 mT, in both static limits, for seven configurations (0–7 without 3; Table S2). For cutoffs of 2a and 3a the bath of radius 5.5a is unchanged and $T_2$ is compared with the 4a result configuration by configuration. A cutoff of 5a needs a bath radius of 6.5a, which changes the random configuration itself, so it is compared with a 4a calculation in the same bath. Going from 4a to 5a changes $T_2$ by at most 0.5 %, except for the static mean field at 30 mT (1.7 %), which converges most slowly. This is smaller than the statistical errors of Table I, and 4a is used throughout.

TABLE S2. Pair-cutoff test: $T_2(r_c)/T_2(4a)$ for $^{14}N$ at 1 ppm (seven configurations; static mean field averaged over two samples per configuration; errors from a bootstrap over configurations, paired with 4a). The 5a column is relative to 4a in the same bath of radius 6.5a.

| B (mT) | treatment | 2a | 3a | 5a |
|---|---|---|---|---|
| 2.5 | no mean field | 1.036 ± 0.009 | 1.007 ± 0.002 | 0.997 ± 0.001 |
| 2.5 | static mean field | 1.076 ± 0.027 | 1.015 ± 0.005 | 0.995 ± 0.001 |
| 30 | no mean field | 1.049 ± 0.014 | 1.009 ± 0.002 | 0.996 ± 0.001 |
| 30 | static mean field | 1.98 ± 0.40 | 1.063 ± 0.043 | 0.983 ± 0.013 |

## APPENDIX S6: CONCENTRATION SCALING

Results are quoted as $T_2 \cdot [N]$ on the assumption $T_2 \propto 1/[N]$. In the model this scaling is not automatic: the dipolar couplings, the Overhauser field and the golden-rule rates scale with [N], whereas the hyperfine and Zeeman energies do not. It holds as long as P1–P1 couplings (about 9 kHz at 1 ppm) are small compared with the separations of the hyperfine-resolved levels, so that pair flip-flops are either resonant or strongly detuned. We tested it at 0.3, 3 and 10 ppm for 23 configurations. The random positions generated with a given seed are exact scaled copies of the 1 ppm configuration, and the Jahn-Teller axes, the initial states and the static samples are the same, so the ratio to 1 ppm can be evaluated configuration by configuration. Table S3 and Fig. S3 show the result: $T_2 \cdot [N]$ is independent of concentration in all treatments, the pair-flip-flop result within 2 % and the independent-jump result within 4 % over a factor of 30 in [N]. The ratio of the two dynamical results, 1.07–1.31, depends on the field but not on the concentration.

TABLE S3. $T_2 \cdot [N]$ (μs·ppm) versus P1 concentration ($^{14}N$, 23 configurations; static mean field with two samples per configuration). The last column is the pair-flip-flop result relative to 1 ppm for the same configurations.

| B (mT) | [N] (ppm) | no mean field | indep. jumps | pair flip-flops | static | pair / 1 ppm |
|---|---|---|---|---|---|---|
| 2.5 | 0.3 | 76 ± 5 | 104 ± 8 | 113 ± 9 | 123 ± 9 | 1.013 ± 0.014 |
| 2.5 | 1 | 76 ± 5 | 103 ± 7 | 112 ± 8 | 122 ± 9 | 1.000 ± 0.000 |
| 2.5 | 3 | 76 ± 5 | 103 ± 7 | 111 ± 9 | 121 ± 9 | 0.995 ± 0.011 |
| 2.5 | 10 | 76 ± 5 | 103 ± 8 | 111 ± 9 | 120 ± 9 | 0.993 ± 0.010 |
| 10 | 0.3 | 70 ± 4 | 140 ± 9 | 170 ± 11 | 235 ± 20 | 1.017 ± 0.016 |
| 10 | 1 | 70 ± 4 | 138 ± 9 | 167 ± 11 | 235 ± 20 | 1.000 ± 0.000 |
| 10 | 3 | 70 ± 3 | 137 ± 9 | 168 ± 11 | 234 ± 20 | 1.005 ± 0.012 |
| 10 | 10 | 70 ± 4 | 135 ± 9 | 163 ± 11 | 227 ± 19 | 0.978 ± 0.014 |
| 30 | 0.3 | 68 ± 3 | 147 ± 9 | 185 ± 11 | 264 ± 17 | 0.983 ± 0.011 |
| 30 | 1 | 68 ± 3 | 147 ± 9 | 188 ± 12 | 264 ± 16 | 1.000 ± 0.000 |
| 30 | 3 | 68 ± 3 | 147 ± 9 | 186 ± 12 | 263 ± 16 | 0.989 ± 0.015 |
| 30 | 10 | 68 ± 3 | 142 ± 9 | 185 ± 11 | 260 ± 16 | 0.982 ± 0.014 |
| 50 | 0.3 | 68 ± 3 | 149 ± 9 | 190 ± 11 | 266 ± 15 | 1.000 ± 0.007 |
| 50 | 1 | 68 ± 3 | 146 ± 10 | 190 ± 12 | 266 ± 16 | 1.000 ± 0.000 |
| 50 | 3 | 68 ± 3 | 148 ± 10 | 192 ± 11 | 265 ± 15 | 1.009 ± 0.009 |
| 50 | 10 | 68 ± 3 | 144 ± 9 | 186 ± 12 | 262 ± 16 | 0.976 ± 0.011 |

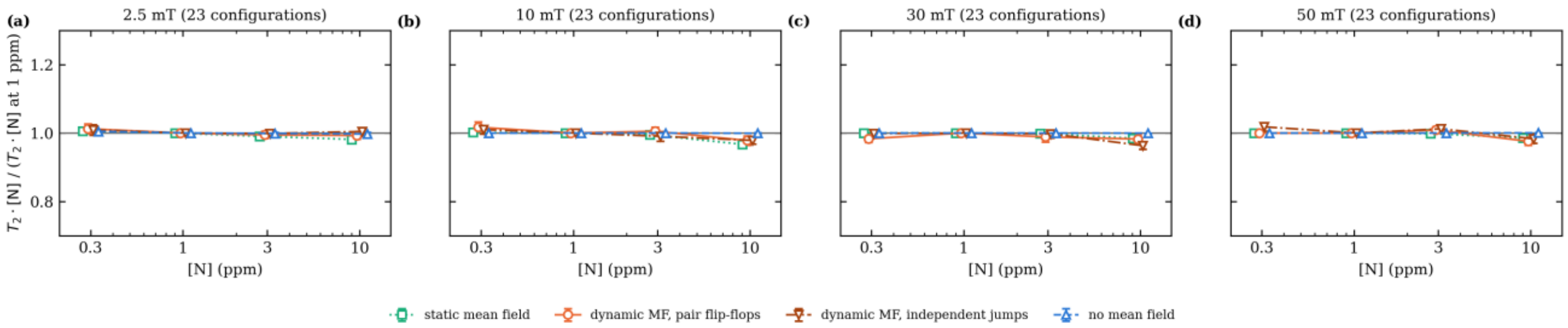

FIG. S3. $T_2 \cdot [N]$ versus P1 concentration, relative to its 1 ppm value for the same configurations, at 2.5, 10, 30 and 50 mT (panels (a), (b), (c), (d)) for the treatments of Fig. 3 ($^{14}N$, 23 configurations; points displaced horizontally for clarity). Error bars are from a bootstrap over configurations paired with 1 ppm; absolute values are given in Table S3. A constant ratio of 1 corresponds to $T_2 \propto 1/[N]$.

## APPENDIX S7: ECHO DECAYS

The $T_2$ values of the main text are 1/e times of configuration-averaged echo decays. For completeness, the decays themselves are shown here. Figure S4 shows all treatments at all four fields; the 2.5 and 30 mT panels repeat Fig. 3 of the main text. The dynamical decays are close to the mean-field-free one at short times, where only the strongly coupled close pairs contribute and these are insensitive to the outer field, and approach the static one at long times, where the slowly changing part of the Overhauser field dominates; the pair flip-flops stay closer to the static decay than the independent jumps. Figure S5 shows the decays for the rate scales $\lambda = 0.3$, 1 and 3 used in Sec. III D (independent jumps), which lie between the two limits and move monotonically with $\lambda$. Figure S6 shows the pair-flip-flop decays at 0.3, 1, 3 and 10 ppm plotted against $2\tau \cdot [N]$; they fall on top of each other, which is the scaling of Sec. S6 seen in the full decay rather than in its 1/e point alone.

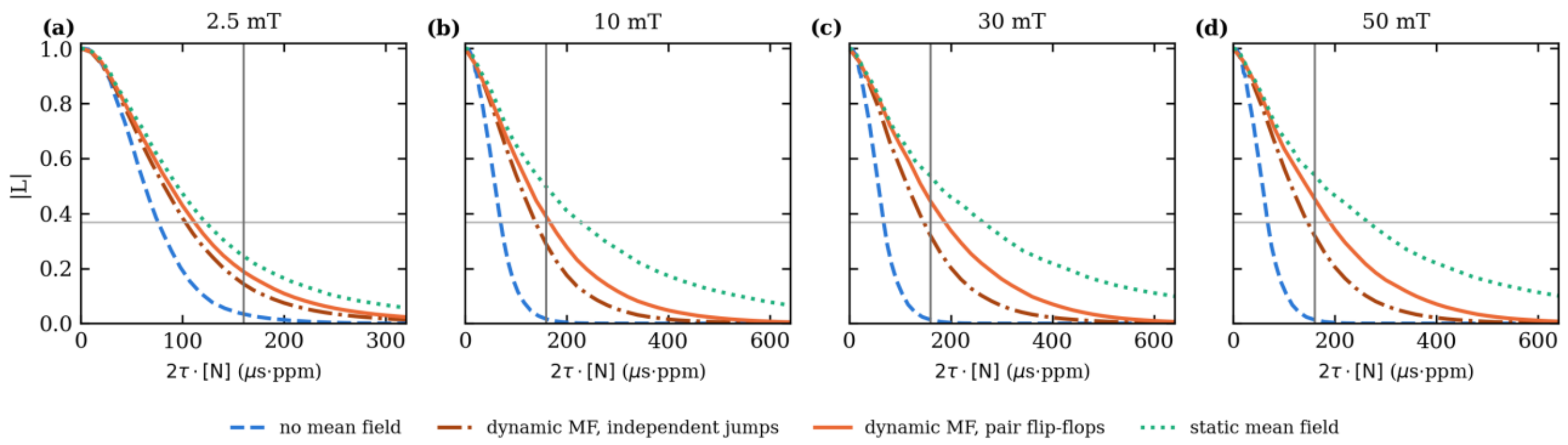

FIG. S4. Configuration-averaged Hahn-echo decays of an $NV^{-}$ centre in a $^{14}N$ P1 bath at 2.5, 10, 30 and 50 mT (23 configurations): no mean field (dashed), dynamical mean field at the golden-rule rates with pair flip-flops (solid) and independent jumps (dash-dotted), and static mean field (dotted). The vertical line marks $T_2 \cdot [N] = 160$ μs·ppm [2] and the horizontal line 1/e.

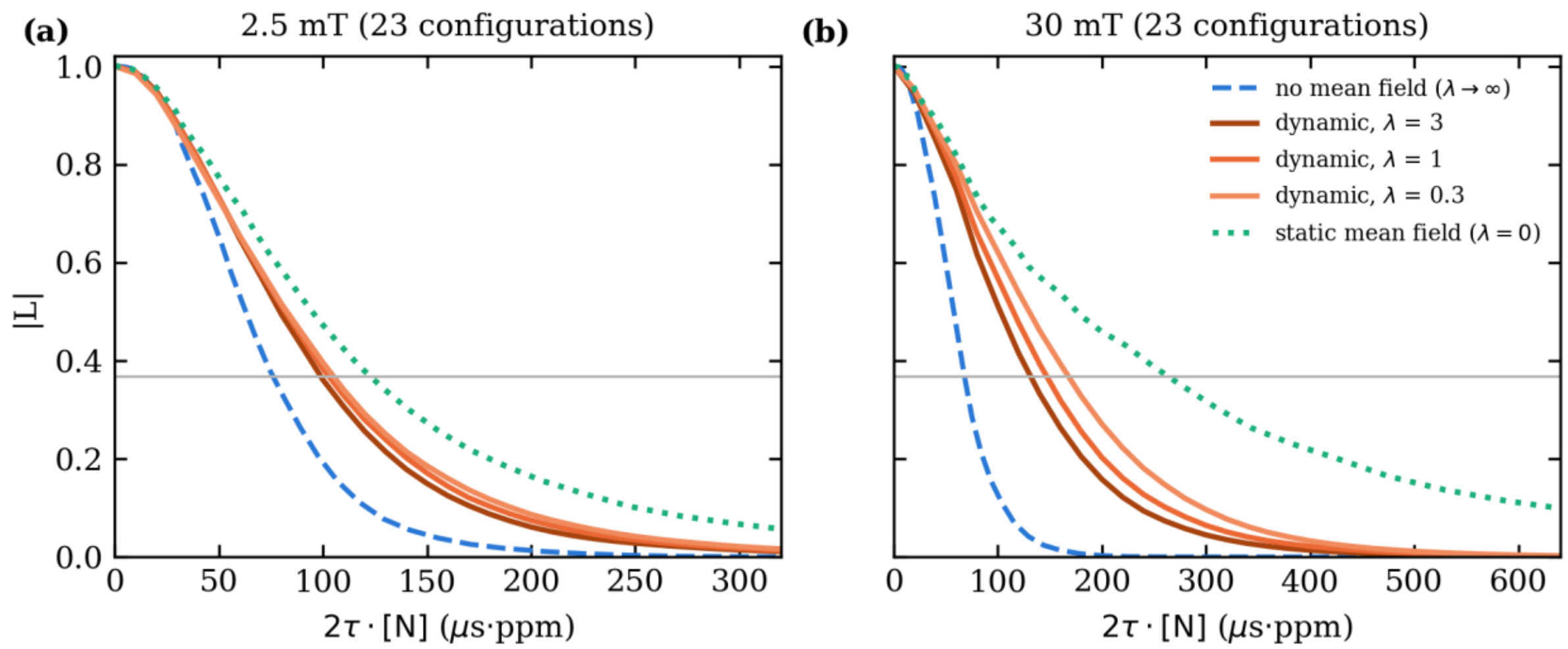

FIG. S5. Echo decays for the rate scales λ = 0.3, 1 and 3 (independent jumps; solid, light to dark) together with the mean-field-free (dashed) and static (dotted) limits at (a) 2.5 mT and (b) 30 mT, for the 23 configurations of the sweep.

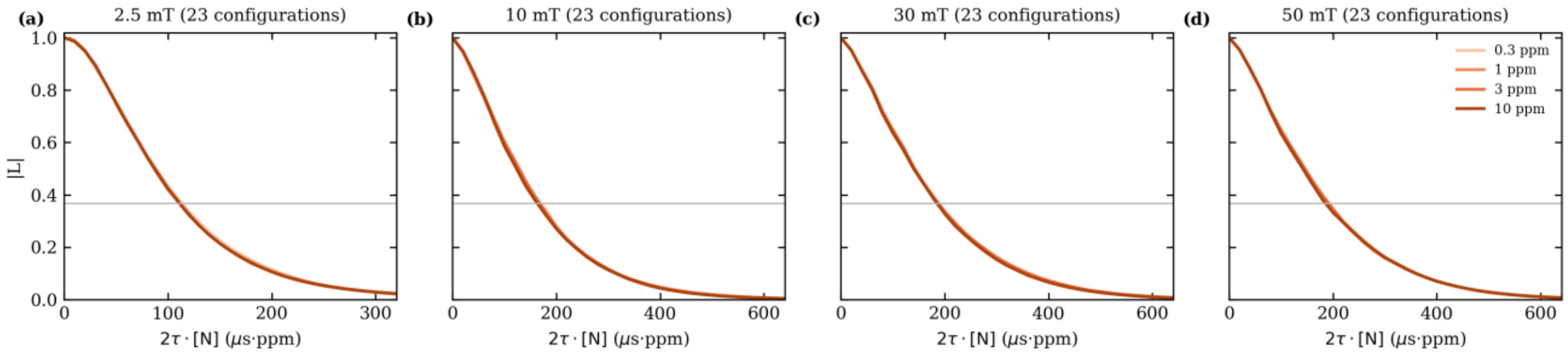

FIG. S6. Dynamical-mean-field echo decays (pair flip-flops) at 0.3, 1, 3 and 10 ppm (light to dark) plotted against 2τ·[N] at 2.5, 10, 30 and 50 mT ($^{14}N$, 23 configurations). At 2.5 mT and 3 and 10 ppm the CCE2 coherence shows isolated single-time-point excursions, a known artefact of the expansion where a subcluster contribution passes close to zero — here at zeros of the intra-P1 hyperfine modulation, whose frequencies do not scale with [N]. The curves are therefore shown as the running minimum of |L|, the quantity from which $T_2$ is read throughout; $T_2$ obtained instead from a median-filtered decay differs by less than 2 %.

## APPENDIX S8: $T_2$ FROM A STRETCHED-EXPONENTIAL FIT

The main text reads $T_2$ as the 1/e time of the running minimum of the configuration-averaged |L|. This estimator needs no model of the decay shape. Here we compare it with the more common fit to a stretched exponential, $|L| = \exp[-(2\tau/T_2)^p]$. We fit $\ln(-\ln|L|)$ linearly in $\ln 2\tau$ over the points with $0.1 < |L| < 0.95$, for the same configurations as Table I. The errors come from the same bootstrap over configurations, and the ratio of the two estimates is paired within each bootstrap sample.

Table S4 shows the result. The two estimates agree within 2 % for the pair flip-flops and within 4 % for all treatments, well inside the statistical errors. With the fitted values, the field dependence of the pair flip-flops, $T_2(30\ \text{mT})/T_2(2.5\ \text{mT})$, is 1.66 instead of 1.69. A narrower fit window ($0.2 < |L| < 0.9$) or a wider one ($0.05 < |L| < 0.98$) changes the fitted pair-flip-flop $T_2$ by less than 5 %. The stretching exponent itself carries information. The mean-field-free decay is close to Gaussian ($p \approx 2.0$). The static mean field gives $p \approx 1.1$ at 30 mT, close to the $p \approx 1$ that Schätzle et al. found for bulk diamond [8]. The pair flip-flops lie in between, with $p \approx 1.3$ at 30 mT.

TABLE S4. $T_2 \cdot [N]$ (μs·ppm) from the 1/e time (main text) and from a stretched-exponential fit, with the fitted exponent p ($^{14}$N, 1 ppm, 23 configurations). The last column is the ratio of the two estimates, paired within the bootstrap.

| B (mT) | treatment | 1/e time | fit | p | fit / 1/e |
|---|---|---|---|---|---|
| 2.5 | no mean field | 76 ± 5 | 77 ± 5 | 2.08 ± 0.11 | 1.011 ± 0.012 |
| 2.5 | indep. jumps | 103 ± 8 | 105 ± 8 | 1.65 ± 0.08 | 1.018 ± 0.014 |
| 2.5 | pair flip-flops | 112 ± 8 | 114 ± 10 | 1.54 ± 0.08 | 1.021 ± 0.017 |
| 2.5 | static | 123 ± 10 | 128 ± 11 | 1.53 ± 0.08 | 1.039 ± 0.018 |
| 10 | no mean field | 70 ± 3 | 70 ± 4 | 2.05 ± 0.10 | 1.002 ± 0.013 |
| 10 | indep. jumps | 138 ± 9 | 139 ± 9 | 1.48 ± 0.07 | 1.001 ± 0.010 |
| 10 | pair flip-flops | 167 ± 11 | 167 ± 12 | 1.34 ± 0.05 | 1.000 ± 0.010 |
| 10 | static | 228 ± 18 | 231 ± 17 | 1.17 ± 0.05 | 1.014 ± 0.020 |
| 30 | no mean field | 68 ± 3 | 69 ± 4 | 2.03 ± 0.11 | 1.013 ± 0.014 |
| 30 | indep. jumps | 147 ± 9 | 147 ± 10 | 1.49 ± 0.07 | 1.000 ± 0.009 |
| 30 | pair flip-flops | 188 ± 12 | 190 ± 12 | 1.32 ± 0.06 | 1.005 ± 0.008 |
| 30 | static | 264 ± 16 | 260 ± 17 | 1.12 ± 0.05 | 0.984 ± 0.019 |
| 50 | no mean field | 68 ± 3 | 69 ± 4 | 2.03 ± 0.11 | 1.014 ± 0.014 |
| 50 | indep. jumps | 146 ± 9 | 147 ± 10 | 1.48 ± 0.06 | 1.005 ± 0.010 |
| 50 | pair flip-flops | 190 ± 11 | 189 ± 12 | 1.32 ± 0.05 | 0.994 ± 0.009 |
| 50 | static | 266 ± 16 | 261 ± 18 | 1.12 ± 0.05 | 0.982 ± 0.020 |

## APPENDIX S9: SENSITIVITY OF THE PAIR FLIP-FLOPS

The λ sweep of Sec. III D uses independent jumps. Here we test the pair flip-flops directly, for seven configurations (0–7 without 3), at 2.5 and 30 mT. We change one setting at a time. The rate scale λ takes the values 0.3 and 3. The width of the Gaussian line shape in Eq. (3) takes the values $\sigma = |V|$ and $4|V|$ instead of $2|V|$; the rejection bound of the kinetic Monte Carlo is rebuilt with the same width. All else stays as in the main calculation: the bath, the screening of the pairs, the initial states and the random numbers of the trajectories. A rerun with the original settings reproduces the main coherences exactly. $T_2$ is therefore compared with the main result configuration by configuration, and the errors come from a bootstrap over configurations paired with the main result.

Table S5 shows the result. At 30 mT a wider line shape and faster rates both shorten $T_2$, because they let more off-resonant pairs flip and make the Overhauser field less static. The changes stay within about 10 %. At 2.5 mT none of the four changes exceeds the statistical error. There the hyperfine substructure decides which pairs are resonant, and the rates and widths matter little.

TABLE S5. Sensitivity of the pair-flip-flop result: $T_2$(changed)/$T_2$(main) for $^{14}N$ at 1 ppm (seven configurations; main calculation $\sigma = 2|V|$, $\lambda = 1$). Errors are from a bootstrap over configurations paired with the main result.

| **changed setting** | **2.5 mT** | **30 mT** |
|---|---|---|
| $\sigma = \|V\|$ | 1.005 ± 0.019 | 1.058 ± 0.031 |
| $\sigma = 4\|V\|$ | 0.988 ± 0.023 | 0.929 ± 0.013 |
| $\lambda = 0.3$ | 0.995 ± 0.033 | 1.059 ± 0.028 |
| $\lambda = 3$ | 0.993 ± 0.021 | 0.896 ± 0.036 |